\documentclass[twocolumn]{aastex631}

\usepackage{amsmath}

\usepackage[T1]{fontenc}
\usepackage[utf8x]{inputenc}

\usepackage{graphicx}
\usepackage{dcolumn}
\usepackage{verbatim}

\usepackage{bm}
\usepackage{breakurl}
\usepackage{lmodern}

\usepackage{tikz}
\usepackage{pgfplots}
\pgfplotsset{compat=newest,every axis plot/.append style={line width=1pt}}

\usepackage{tabularx}

\definecolor{lightgray}{gray}{0.9}
\definecolor{Amber}{rgb}{1.0, 0.75, 0.0}
\definecolor{blizzardblue}{rgb}{0.67, 0.9, 0.93}
\definecolor{Amber_mod}{rgb}{1.0, 0.35, 0.0}

\renewcommand{\arraystretch}{1.5}

\begin{document}

\reportnum{YITP-21-109, IPMU21-0064} 

\title{Establishing the Non-Primordial Origin of Black Hole-Neutron Star Mergers}

\author{Misao Sasaki}
\email{misao.sasaki@ipmu.jp}
\affiliation{Kavli Institute for the Physics and Mathematics of the Universe (WPI), UTIAS, The University of Tokyo, Chiba 277-8583, Japan}
\affiliation{Center for Gravitational Physics, Yukawa Institute for Theoretical Physics, Kyoto University, Kyoto 606-8502, Japan}
\affiliation{Leung Center for Cosmology and Particle Astrophysics, National Taiwan University, Taipei 10617}

\author{Volodymyr Takhistov}
\email{volodymyr.takhistov@ipmu.jp}
\affiliation{Kavli Institute for the Physics and Mathematics of the Universe (WPI), UTIAS, The University of Tokyo, Chiba 277-8583, Japan}

\author[0000-0002-8496-5859]{Valeri Vardanyan}
\email{valeri.vardanyan@ipmu.jp}
\affiliation{Kavli Institute for the Physics and Mathematics of the Universe (WPI), UTIAS, The University of Tokyo, Chiba 277-8583, Japan}

\author{Ying-li Zhang}
\email{yingli@tongji.edu.cn}
\affiliation{School of Physics Science and Engineering, Tongji University, Shanghai 200092, China}
\affiliation{Institute for Advanced Study of Tongji University, Shanghai 200092, China}
\affiliation{Kavli Institute for the Physics and Mathematics of the Universe (WPI), UTIAS, The University of Tokyo, Chiba 277-8583, Japan}
\affiliation{Center for Gravitation and Cosmology, YangZhou University, Yangzhou 225009, China}


\begin{abstract}
Primordial black holes (PBHs) from the early Universe constitute an attractive dark matter candidate. First detections of black hole-neutron star (BH-NS) candidate gravitational wave events by the LIGO/Virgo collaboration, GW200105 and GW200115, already prompted speculations about non-astrophysical origin. We analyze, for the first time, the total volumetric merger rates of PBH-NS binaries formed via two-body gravitational scattering, finding them to be subdominant to the astrophysical BH-NS rates. In contrast to binary black holes, a significant fraction of which can be of primordial origin, either formed in dark matter halos or in the early Universe, PBH-NS rates cannot be significantly enhanced by contributions preceding star formation. Our findings imply that the identified BH-NS events are of astrophysical origin, even when PBH-PBH events significantly contribute to the GW observations.
\end{abstract}


\section{Introduction} 

The initial breakthrough discovery of gravitational waves (GWs)~\cite{LIGOScientific:2016aoc} opened a new window for exploring astronomical, cosmological as well as particle physics phenomena. Dozens of compact binary merger sources have already been observed by the LIGO/Virgo collaboration (LVC). The vast majority of these events are binary black holes (BH-BH) with components in the $\sim 10 - 100 M_\odot$ mass-range \cite{LIGOScientific:2020ibl}. While a variety of conventional astrophysical stellar evolution channels could contribute to such events (see e.g. \cite{Mandel:2018hfr,Mandel:2021smh} for reviews), comprehensive understanding of their origin is still lacking and could be connected with central puzzles of modern physics, such as the nature of dark matter (DM).

Intriguingly, GW observations are consistent with mergers of primordial black holes (PBHs) formed in the early Universe prior to galaxy and star formation that can contribute to the DM abundance~(e.g.~\cite{Zeldovich:1967,Hawking:1971ei,Carr:1974nx,Cotner:2019ykd,Cotner:2018vug,Sasaki:2018dmp,Kusenko:2020pcg}). Depending on the formation mechanism PBHs can span many orders of magnitude in mass. Binary PBH mergers are able to account for GW observations, such as the GW190521 event with a total merger mass of $\sim 150 M_{\odot}$ lying in the pair-instability supernova mass gap~\cite{LIGOScientific:2020iuh}, which have challenged conventional astrophysical interpretations. In the mass ranges relevant for the current GW detectors PBHs could contribute a sizable fraction of the DM energy density $f_{\rm PBH} = \Omega_{\rm PBH}/\Omega_{\rm DM}$~(e.g.~\cite{Ali-Haimoud:2016mbv,Lu:2020bmd,Serpico:2020ehh,Takhistov:2021aqx}), with GW data suggesting $f_{\rm PBH} \lesssim \mathcal{O}(10^{-3})$~(e.g.~\cite{Bird:2016dcv,Clesse:2016vqa,Sasaki:2016jop,Franciolini:2021tla}), although uncertainties exist. The quest for identifying the origin of BH mergers is being advanced across several directions~(e.g.~\cite{Hutsi:2020sol,DeLuca:2021wjr,Raccanelli:2016cud,Canas-Herrera:2021qxs,Canas-Herrera:2019npr,Mukherjee:2021ags}).  

Besides the binary BH GW events, detection of binary neutron star (NS-NS) mergers in GWs as well as electromagnetic signatures have spearheaded the investigations in multi-messenger astronomy~\cite{LIGOScientific:2017vwq,LIGOScientific:2017ync}. Without identification of clear electromagnetic counterpart signals or sufficient sensitivity to higher
order tidal deformability effects, distinguishing between a solar-mass BH and a NS is difficult. While solar-mass BHs are not expected from conventional stellar evolution, they can readily appear either as PBHs or ``transmuted'' BHs from small sub-solar mass PBHs (or particles) constituting DM being captured and devouring NSs~(e.g.~\cite{Capela:2013yf,Fuller:2017uyd,Bramante:2017ulk,Takhistov:2017bpt,Takhistov:2017nmt}), leading to alternative interpretations of the detected NS merger events~(e.g.~\cite{Kouvaris:2018wnh,Tsai:2020hpi,Takhistov:2020vxs,Dasgupta:2020mqg}).

Recently, LVC has reported first identified BH-NS binary GW events GW200105 and GW200115, with component masses of ($8.9 \substack{+1.2 \\ -1.5} M_\odot, 1.9 \substack{+0.3 \\ -0.2} M_\odot$) and ($5.7 \substack{+1.8 \\ -2.1} M_\odot, 1.5 \substack{+0.7 \\ -0.3} M_\odot$), respectively~\cite{LIGOScientific:2021qlt}.  In addition to the BH-BH and NS-NS mergers, BH-NS events constitute another major class of mergers and carry significant implications for multimessenger observations~\cite{Ruiz:2021gsv}.
While the detected events are consistent with stellar evolution formation channels~\cite{Broekgaarden:2021hlu}, speculations about possible PBH origin, considering that NSs are mis-identified solar-mass BHs and detected events correspond to unequal mass PBH-PBH mergers, have already been put forth~\cite{Wang:2021iwp}. As the number of detected events significantly accumulates in the upcoming future and given their possible implications for fundamental physics, understanding their origin is a pressing matter. An essential ingredient for understanding the role of PBHs and DM in the context of BH-NS events are contributions of PBH-NS mergers, thus far not comprehensively explored.

In this work we analyze, for the first time, the expected average merger rates of PBH-NS events formed in galaxies, conservatively considering that NSs have been properly identified.  Unlike astrophysical BH-NS systems, purely stellar evolution formation channels are not available for PBH-NS binaries and we focus on their dynamical assembly. As we discuss, one of the essential differences with PBH-PBH binaries is that PBH-NS binaries must have been formed in the late Universe, well after the onset of star formation.

\section{PBH-NS Merger Rates}

In order to find the total PBH-NS merger rate, we first consider an isolated galaxy and then discuss contributions from galaxy populations. 

PBH-NS binaries are formed in galaxies via 2-body scattering involving GW emission. We leave more complicated formation channels, such as three-body encounters, for future simulations. Some of such effects have been studied in the context of binary black hole mergers \cite{Kritos:2020wcl}. We stress that the formation channel considered by us is the simplest, relying on minimal assumptions.

Upon approach of a PBH on a hyperbolic orbit to a NS within a critical impact parameter the gravitational wave emission exceeds the initial kinetic energy and leads to formation of a bound PBH-NS system. The capture cross-section for components of masses $m_1$ and $m_2$ is given by~\cite{Quinlan:1989,Mouri:2002mc}
\begin{align}\label{eq:cross_section}
    \sigma = 2\pi \left(\frac{85\pi}{6\sqrt{2}}\right)^{2/7}G_\mathrm{N}^2M^{12/7}\mu^{2/7}c^{-10/7}v_\mathrm{rel}^{-18/7}~,
\end{align}
where $M = m_1 + m_2$, $\mu = m_1 m_2/M^2$, $c$ is the speed of light, $v_\mathrm{rel}$ is the relative velocity between the components and $G_{\rm N}$ is gravitational constant.

In order to calculate the merger rate we need to establish the density overlap between NS and PBH distributions. For DM, including PBHs and particle DM, we assume the galactic halo profile to be given by the Navarro-Frenk-White (NFW) model~\cite{Navarro:1995iw}
\begin{equation} \label{eq:nfw}
\rho_\mathrm{DM}(r) = \rho_0\left[\frac{r}{R_\mathrm{s}}\left(1 + \frac{r}{R_\mathrm{s}}\right)^2\right]^{-1}~,
\end{equation}
where $\rho_0$ and $R_s$ are the characteristic density and radius of the halo. We estimate that our results are not very sensitive to the details of the considered DM profile.

While PBHs are born in the early Universe, NSs are born from gravitational collapse and supernovae explosions of massive stars. NSs are often born with significant ``natal kick'' velocities, reaching hundreds of km/s (e.g.~\cite{Arzoumanian:2001dv}).~As a benchmark, we focus on the Milky Way (MW) Galactic NS population. Majority of NSs have only been observed as isolated pulsars with ages significantly shorter than that of MW, with significant uncertainties on population and distribution - especially towards the Galactic Center~(e.g.~\cite{Sartore:2010}).
We consider exponential galactic NS distribution~\cite{Paczynski:1990}, well motivated by the star formation history of the
MW~\cite{Sartore:2010},  and employ the following spherically symmetric model
\begin{align} \label{eq:nsdis}
    \rho_\mathrm{NS}(r) = \rho^{0}_\mathrm{NS}e^{-r/R_\mathrm{NS}}~,
\end{align}
where $\rho^{0}_\mathrm{NS}$ and $R_\mathrm{NS}$ are the characteristic density and radius, respectively.

The resulting binary formation rate in a particular galactic halo is given by
\begin{align}\label{eq:rate_integral}
    \mathcal{R}_\mathrm{PBH-NS} = 4 \pi \int_0^{R_\mathrm{vir}} \mathrm{d} r r^2 \frac{\rho_\mathrm{NS}}{m_1}\frac{\rho_\mathrm{PBH}}{m_2}\langle\sigma v_\mathrm{rel}\rangle~,
\end{align}
where $R_\mathrm{vir}$ is the virial radius of the halo and the angle brackets denote averaging over the velocity distribution.

\begin{figure}[tb]
    \centering
    \includegraphics[width=\columnwidth]{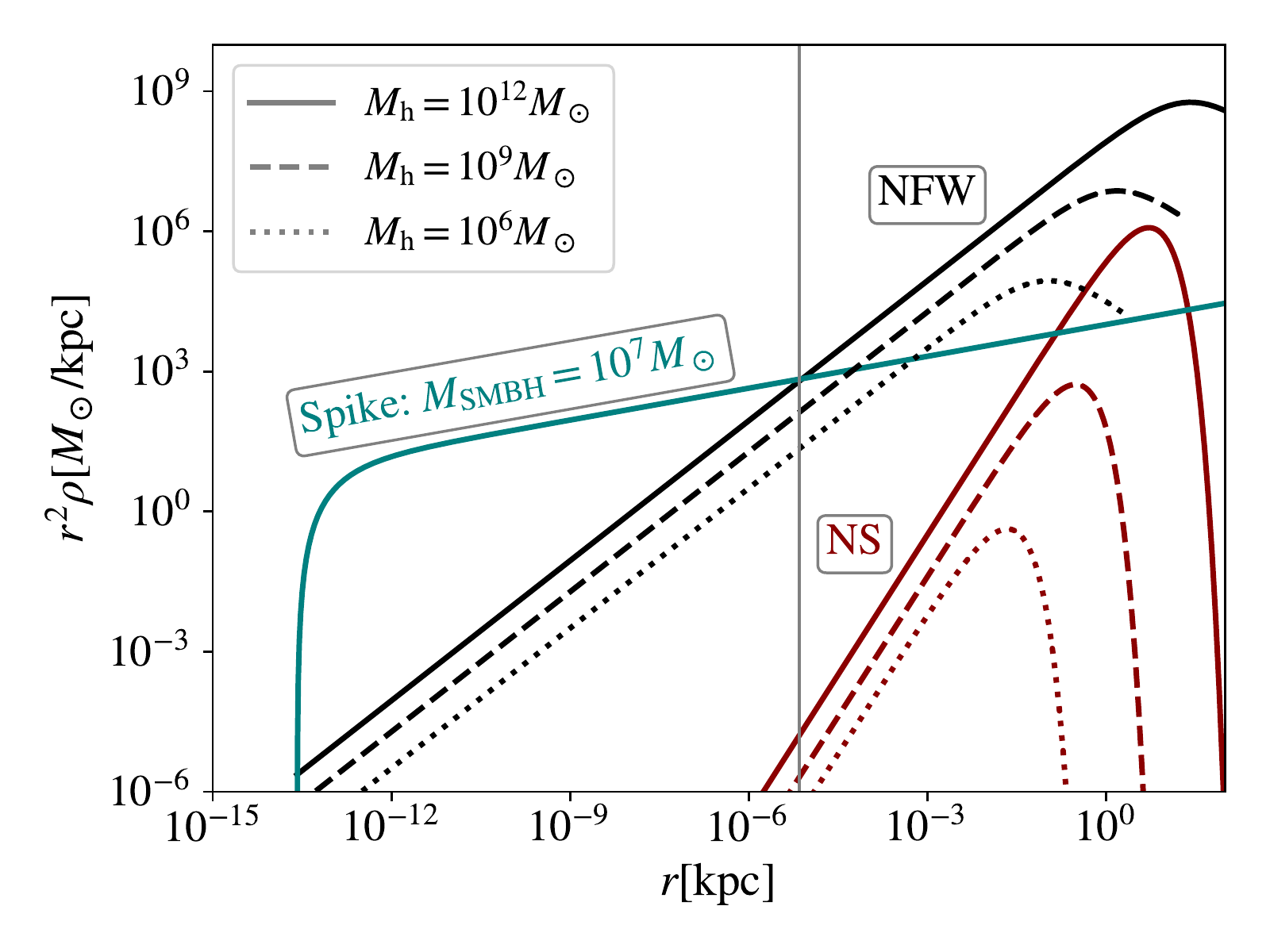}
    \caption{Density profiles of PBH DM halo (NFW, black) and NSs (red) for halos of different masses $M_{\rm h}$, assuming $R_{\rm NS}/R_s = 0.1$ (see the text for definition). Possible DM ``density spike'' enhancement due to a supermassive BH at the galactic center is also shown.}
    \label{fig:density_profiles}
\end{figure}

Typical binaries are expected to have very large eccentricities, therefore their merger times are negligibly small (especially in the case of large halos) \cite{Cholis:2016kqi}. As a result the obtained binary formation rates $\mathcal{R}_\mathrm{PBH-NS}$ can be identified as the merger rates for the given halo.

We are interested in obtaining the merger rates as a function of the halo mass $M_\mathrm{h}$, as well as the cosmological volume averages of these rates. In order to obtain the former we need halo properties as a function of the halo mass. First, we establish a relation between the halo concentration $C\equiv R_\mathrm{vir}/R_\mathrm{s}$ and $M_\mathrm{h}$, which is inferred from N-body simulations. We employ the fitting function provided in \cite{Ludlow:2016ifl} (see their Appendix C). The characteristic NFW density $\rho_0$ of Eq.~\eqref{eq:nfw} is then given by
\begin{align}
    \frac{\rho_0}{\rho_\mathrm{crit}} = \frac{200}{3}\frac{C^3}{g(C)}~,
\end{align}
where $g(C) \equiv \log{(1 + C)} - C/(1 + C)$, and $\rho_\mathrm{crit}$ is the critical density of the universe.
The radius $R_\mathrm{s}$ is given by
\begin{align}
    R_\mathrm{s} = \left(\frac{M_\mathrm{h}}{4\pi g(C)\rho_0}\right)^{1/3}~.
\end{align}

\begin{figure*}[t]
    \centering
    \includegraphics[width=\columnwidth]{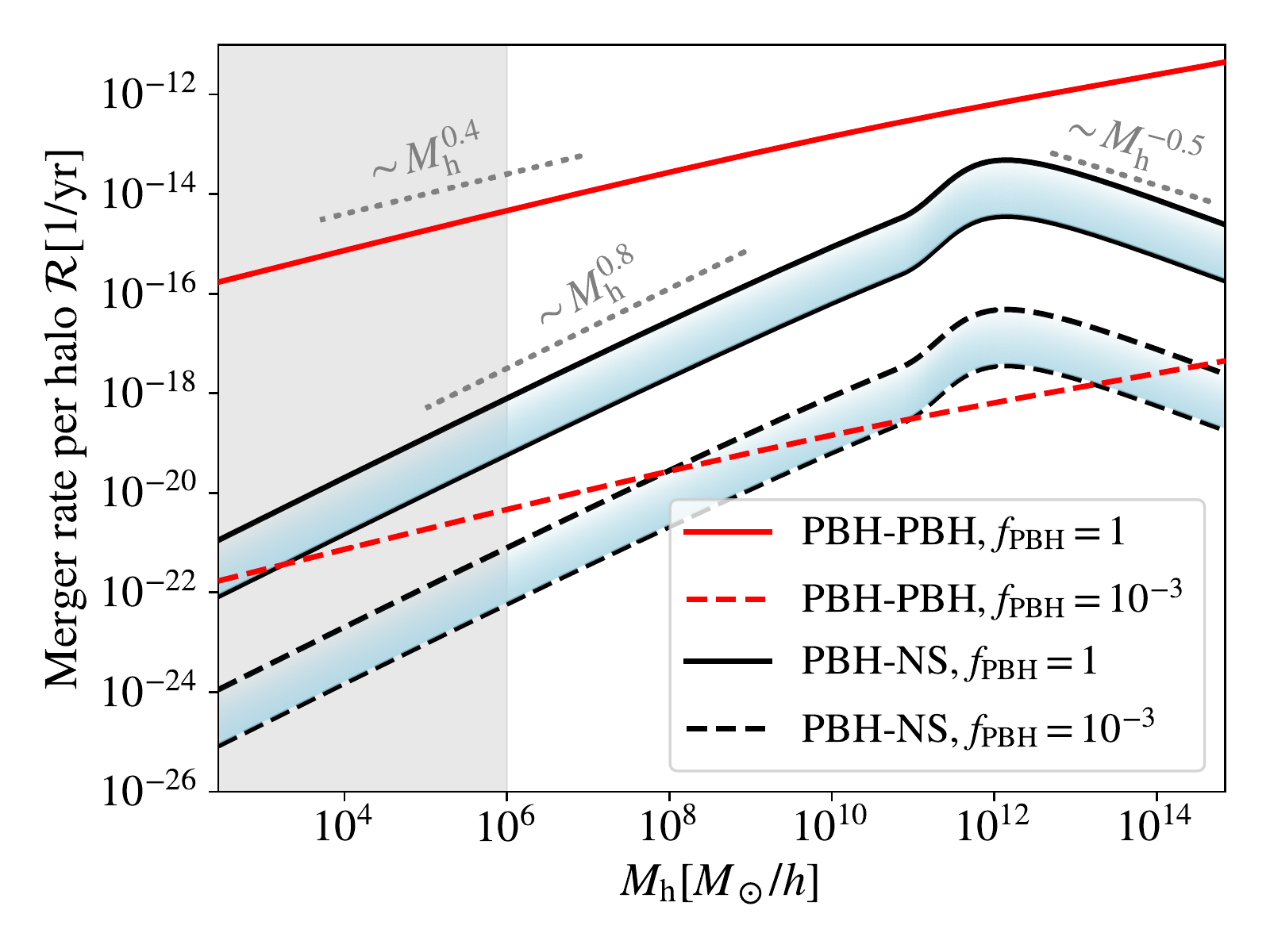}
    \includegraphics[width=\columnwidth]{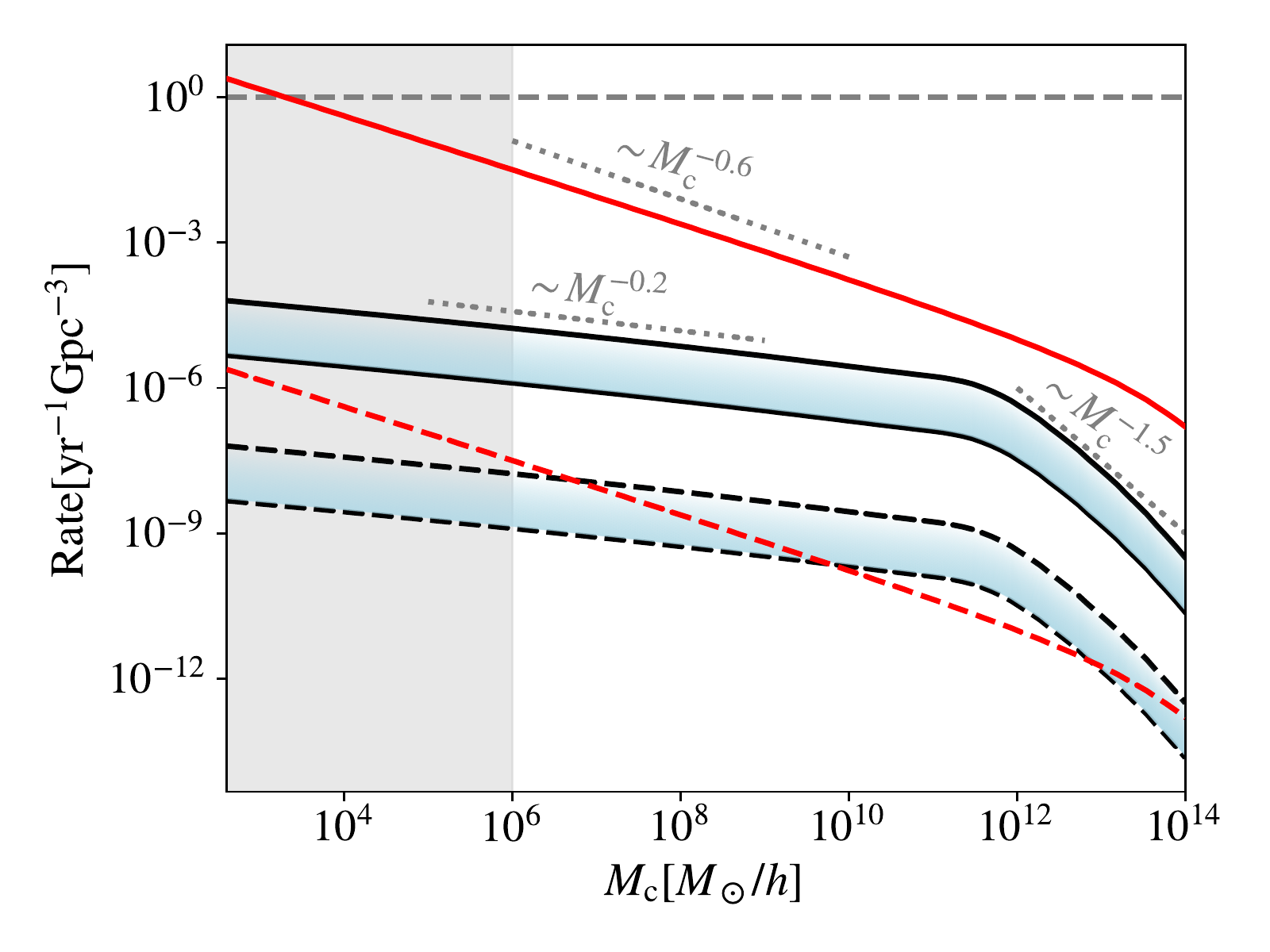}
    \caption{\textit{\textbf{Left:}} PBH-NS and PBH-PBH binary merger rates per halo, as a function of halo mass $M_\mathrm{h}$, for a range of $f_{\rm PBH}$ values. Uncertain extrapolated small halo contributions below $M_h \lesssim 10^6 M_{\odot}$ are shaded in gray. \textit{\textbf{Right:}} PBH-NS and PBH-PBH merger rates integrated over the halo mass function, presented as a function of the smallest contributing halo with mass $M_c$. For comparison, PBH-PBH merger rates in the early-Universe formation channel are independent of halo mass and are estimated to be $\mathcal{O}(1)\mathrm{Gpc}^{-3}\mathrm{yr}^{-1}$ for $f_\mathrm{PBH} = 10^{-3}$ and $\mathcal{O}(10^5)\mathrm{Gpc}^{-3}\mathrm{yr}^{-1}$ for $f_\mathrm{PBH} = 1$ (see, e.g. \cite{Sasaki:2016jop}).}
    \label{fig:rates}
\end{figure*}

Analogously to the DM distribution, two parameters need to be fixed to characterize the NS distribution in Eq.~\eqref{eq:nsdis}. For $R_\mathrm{NS}$ we consider a range of $R_\mathrm{NS}/R_\mathrm{s}$ values. The parameter $\rho^{0}_\mathrm{NS}$ is obtained by normalizing the NS distribution to the estimated number of NSs in a given galaxy determined based on the galaxy's stellar mass and the stellar mass function. For the latter we use the standard Salpeter initial stellar mass-function $\phi(m_\ast)\sim m_\ast^{-2.35}$
and assume that it is independent of time. Time dependence can be easily incorporated into the analysis, but we do not expect it to alter our conclusions. We consider that all the stars in the $[m^\mathrm{min}_\ast, m^\mathrm{max}_\ast] = [8 - 20]M_\odot$ range undergo a supernova explosion leading to a NS remnant. Hence, the number of NSs in a galaxy with a stellar mass $M_\ast$ is given by
\begin{align}
    N_\mathrm{NS}\left(M_\ast\right) = M_\ast\int_{m^\mathrm{min}_\ast}^{m^\mathrm{max}_\ast}\mathrm{d} m_\ast \phi(m_\ast)~,
\end{align}
where $\phi(m_\ast)m_\ast$ is normalized to unity.

The galactic stellar mass $M_\ast$ is extrapolated from the stellar mass-halo mass relation presented in Ref.~\cite{Behroozi_2013}. 
Note that this relation is in principle only valid for central galaxies, but it is sufficient for our purposes as we do not expect a sizeable population of NSs in faint satellites. We should also emphasize that there is a significant uncertainty at the lower mass end of the $M_\ast(M_\mathrm{h})$ relation. As we will see, the smaller halos do not significantly contribute to the PBH-NS rates, and the mentioned scatter is not essential for our main results. In order to provide optimistic estimates, we have chosen the least steep model with $M_\ast\sim M_\mathrm{h}^{1.4}$ for $M_\mathrm{h}\lesssim 10^{12}M_\odot$. In larger halos the relation is almost flat, with $M_\ast\sim M_\mathrm{h}^{0.2}$ for $M_\mathrm{h}\gtrsim 10^{12}M_\odot$.

In Fig.~\ref{fig:density_profiles} we demonstrate the relative distribution of NSs (red) and DM (black) in halos of different masses, assuming $R_\mathrm{NS}/R_\mathrm{s} = 0.1$. Subsequently, we consider the full range  $R_\mathrm{NS}/R_\mathrm{s} \in [0.01, 0.1]$. While the exact value of $R_\mathrm{NS}$ is relevant for the precise value of the merger rate, our overall conclusion is largely insensitive to it.
 
Our estimates for the expected yearly merger rates per halo are presented in the left panel of Fig.~\ref{fig:rates}, where solid and dashed lines correspond to $f_\mathrm{PBH} = 1$ and $f_\mathrm{PBH} = 10^{-3}$, respectively. PBH-NS rates are displayed in black, and the blue gradient-shaded regions correspond to variation in $R_\mathrm{NS}/R_\mathrm{s}$.  

We find that PBH-NS rates significantly decrease with halo mass $M_\mathrm{h}$. This can be understood from the sharp drop of stellar mass for halos with $M_\mathrm{h} \lesssim 10^{12}M_\odot$, which drastically reduces the number of available NSs in such halos. Moreover, our extrapolations of the stellar budget to halos with $M_\mathrm{h} \lesssim 10^{6}M_\odot$ (gray shaded area in Fig.~\ref{fig:rates}) are highly uncertain; such small halos are not expected to be massive enough to ignite star formation. 

With gray dotted lines we also demonstrate in the left panel of  Fig.~\ref{fig:rates} the approximate halo-mass scaling of the rates. The slopes of these lines can be approximated by assuming $\rho_0$ is independent of mass, and that halos are given by a top-hat model (in the case of PBH-NS rates). 

To obtain the total galactic merger rates as relevant for LVC observations we have convoluted the merger rates per halo $\mathcal{R}_\mathrm{PBH-NS}$ with the halo mass-function ${\rm d}n/{\rm d}M_{\rm h}$,
\begin{equation} \label{eq:totrate}
    \mathcal{V}_\mathrm{PBH-NS} = \int_{M_c} \mathcal{R}_\mathrm{PBH-NS} \frac{{\rm d}n}{{\rm d}M_\mathrm{h}} {\rm d}M_\mathrm{h}~,
\end{equation}
where $M_\mathrm{c}$ is the lower cutoff limit for the contributing halos.
Here we have employed the Tinker halo mass-function \cite{Tinker:2008ff}. In order to asses the impact of low-mass halos we have varied $M_\mathrm{c}$ in Eq.~\eqref{eq:totrate}, with the resulting total merger rates as a function of $M_\mathrm{c}$ shown in the right panel of Fig.~\ref{fig:rates}. We also display the scaling of the integrated rates with respect to $M_\mathrm{c}$ in gray dotted lines, which are obtained by convoluting the approximate scaling relations of the left panel with the Press-Schechter mass function $\sim M_\mathrm{h}^{-2}$.  

We observe that the curves plateau for $M_\mathrm{h} \lesssim 10^{12}M_\odot$, which is a manifestation of the fact that small halos do not contribute significantly to the PBH-NS rates. 

The LVC observations inferred a BH-NS merger rate of $45 \substack{+75 \\ -33}$~Gpc$^{-3}$yr$^{-1}$, assuming the detected events are representative of the underlying BH-NS population~\cite{LIGOScientific:2021qlt}. Hence, our results establish that the PBH-NS mergers can contribute only as a strictly subdominant component of the observed BH-NS rates. This conclusion is even stronger considering the more realistic case of constrained PBH abundance $f_\mathrm{PBH} = 10^{-3}$, leading to PBH-NS rates being suppressed by an additional factor of $f_\mathrm{PBH}$. 

\section{Comparison with PBH-PBH Mergers}

To put our results for BH-NS rates into context, we recompute for comparison the late Universe PBH-PBH merger rates using Eq.~\eqref{eq:rate_integral} for 2-body scattering with appropriate substitution of $\rho_{\rm PBH}$ for $\rho_{\rm NS}$. Our results for the yearly PBH-PBH merger rates per halo are presented in Fig.~\ref{fig:rates} (in red). Here solid and dashed lines correspond to $f_\mathrm{PBH} = 1$ and $f_\mathrm{PBH} = 10^{-3}$, respectively.

The solid red curve confirms the results previously obtained by Ref.~\cite{Bird:2016dcv}, where it was shown that a significant contribution to PBH-PBH mergers originates in halos as small as $\sim 10^{3} M_\odot$. This is in contrast to the PBH-NS rates, which strongly depend on stellar budget of halos. The magnitude of PBH-NS merger rates approaches non-negligible fraction of PBH-PBH rates in MW-type halos with $M_\mathrm{h} \sim 10^{12}M_\odot$.
Since the PBH-PBH merger rates scale as $\propto f_{\rm PBH}^2$, for the more realistic case of constrained $f_{\rm PBH} \sim 10^{-3}$ they become heavily suppressed and negligible even in larger halos.

We further compute the average PBH-PBH rates by convoluting the halo rates with the halo mass-function in Eq.~\eqref{eq:totrate}. The resulting merger rates as a function of $M_\mathrm{c}$ are shown in the right panel of Fig.~\ref{fig:rates}. We confirm the conclusion of Ref.~\cite{Bird:2016dcv} (note that in the right panel we display the cumulative integral, not the integrand alone), finding that the average PBH-PBH rate can be $\sim \mathcal{O}(1)  \mathrm{Gpc}^{-3} \mathrm{yr}^{-1}$ if $f_\mathrm{PBH} = 1$ and the contributions from small halos are taken into account. As we saw, this is not the case for PBH-NS rates, as these are suppressed in smaller halos due to steep decline in stellar mass.

PBH-NS mergers follow the stellar evolution and can only form at low redshifts. PBH-PBH mergers, on the other hand, can receive significant contributions not only from the late Universe~(e.g.~\cite{Bird:2016dcv}), but also from the early Universe, before matter-radiation equality~(e.g.~\cite{Nakamura:1997sm,Sasaki:2016jop}). As a result, PBH-PBH mergers can significantly contribute to GW observations even if $f_{\rm PBH} \sim 10^{-3}$, while PBH-NS rates are subdominant regardless of the value of $f_{\rm PBH}$. 

\section{Possible Enhancement Effects}

In our analysis we have neglected several possible nuances that might affect the spatial distribution of PBHs and hence merger rates. We now discuss the expected dominant effects and argue that our conclusion about PBH-NS binaries is robust against them.

It has been suggested that PBHs could be clustered on small scales already in the early Universe and be part of ultra-faint dwarf galaxies \cite{Clesse:2016vqa}. Even though these structures can be relevant for PBH-PBH rates, their stellar content is negligible and hence we don't expect significant contribution to the PBH-NS rates. Further, we do not expect that any local environments of NS and DM overdensities can significantly modify the total volumetric galactic merger rates.

Another enhancement effect could be related to DM ``density spikes'' that might form due to accretion by the central galactic supermassive black holes~\cite{Gondolo:1999ef} (see Fig.~\ref{fig:density_profiles}). This can demonstrably influence the PBH-PBH merger rates~\cite{Nishikawa:2017chy}. Following Ref.~\cite{Nishikawa:2017chy}, we have estimated the PBH-NS rates due to the DM spike and have found the contribution to be negligible, being $\mathcal{O}(10^{-20})\text{yr}^{-1}$ per spike, even under the most optimistic assumptions about the spike properties. This can be understood by noticing that the spike spans a very limited volume. While this limitation is overcome in the case of PBH-PBH binary formation due to two factors of DM density in the merger rate integrals (see Eq.~\eqref{eq:rate_integral}), it leads to a negligible effect for the PBH-NS merger rates as these include only a single power of DM density. 

The merger rates could also be enhanced if PBHs have increased concentration towards centers of galaxies. This effect could result from a variety of processes, such as multitude of local gravitational encounters or dynamical friction when PBHs constitute a sub-dominant DM component. In particular, the equilibrium states of multi-component gravitational halos are expected to reach kinetic energy equipartition of the individual species due to encounters (see, e.g.~\cite{galactic_dynamics_2008}). When $f_\mathrm{PBH} < 1$ PBHs will tend to concentrate at the halo center. PBHs might also slow down due to dynamical friction from gravitational attraction of the underlying cold DM, also leading to their concentration at the halo center. 

In case complete mass-segretation between PBHs and other DM components does occur, we expect all of the PBHs in the halo to be clustered within a sphere of radius $R_\mathrm{PBH}$. Introducing a concentration parameter for the PBH halo as $C_\mathrm{PBH} \equiv R_\mathrm{PBH}/R_\mathrm{s} = CR_\mathrm{PBH}/R_\mathrm{h}$ we have 
    $4\pi \rho_\mathrm{s} R_\mathrm{s}^3g(C_\mathrm{PBH}) = f_\mathrm{PBH}M_\mathrm{h}$.
Assuming $C_\mathrm{PBH} \ll 1$ we obtain   $R_\mathrm{PBH} \sim \sqrt{f_\mathrm{PBH}}R_\mathrm{h}$. The characteristic PBH 
velocities are given by $v_\mathrm{PBH}^2 = G_\mathrm{N}f_\mathrm{PBH}M_\mathrm{h}/R_\mathrm{PBH}$. As a result we find that after PBHs concentrate at the halo center their velocities are reduced with respect to the virial velocity of the halo $v_\mathrm{PBH} \sim f_\mathrm{PBH}^{1/4}v_\mathrm{vir}$.
The smaller PBH fractions lead to stronger velocity suppression. 

Using Eqs.~(\ref{eq:cross_section}) and (\ref{eq:rate_integral}), the above idealistic considerations suggest the PBH-NS merger rate to be modified as $\mathcal{O}(1)f_\mathrm{PBH}^{-39/28}\mathcal{R}_\mathrm{PBH-NS}$. For $f_\mathrm{PBH} = 10^{-3}$ this leads to $\mathcal{O}(10^4)$ enhancement. Note that the resulting rate could be even higher than the expected rate in the case of $f_\mathrm{PBH} = 1$. While significant, this enhancement is still far insufficient for matching the expected merger rates to observations (see Fig.~\ref{fig:rates}). More importantly, the time-scales of the mentioned processes are expected to be longer than the age of the Universe in typical halos dominating the PBH-NS rates, therefore the anticipated enhancement would be much weaker than the estimate above.

\section{Concluding remarks}

GW observations allow for new unprecedented tests of fundamental physics.
With first detection of BH-NS mergers and their potential connection to PBHs and the DM, understanding the origin of such events is a central topic of exploration.
We computed, for the first time, the total volumetric PBH-NS merger rates, finding such contributions to be significantly sub-dominant to astrophysical BH-NS rates. Analogously, we recomputed PBH-PBH formation rates in the late Universe, confirming the literature results. While their late Universe rates could be suppressed, PBH-PBH mergers, unlike the PBH-NS ones, can receive significant contributions from the early Universe prior to star formation. As a result, BH-NS binaries are not sensitive probes of PBH dark matter. They also will not significantly contribute to multimessenger observations. The above carries important fundamental implications for a large class of mergers. Namely, not only the detected BH-NS events are of astrophysical origin, but this is the case even when PBH-PBH events account for LVC BH-BH observations.

Additionally, our conclusion regarding the observed BH-NS mergers being of astrophysical origin has further broader implications for BH-BH systems as well. Once a given astrophysical stellar evolution formation channel predicts the expected BH-NS abundance, the same channel also would have a unique prediction for the astrophysical BH-BH rates. For instance, once the observational uncertainties in BH-NS rates are sufficiently reduced, population-synthesis-based analyses will be able to narrow down on the star formation properties, such as metallicities. The latter would determine the BH-BH rates, hence also constraining the scope of contributions from PBHs.

\begin{acknowledgments}
We thank Metin Ata, Omar Contigiani and Masahiro Takada for useful comments. This work is supported in part by JSPS KAKENHI grants (19H01895,  20H04727,  20H05853).
M.S., V.T. and V.V. are also supported by the WPI Research Center Initiative, MEXT, Japan. V.V. is supported by JSPS KAKENHI grant (20K22348). Y.Z. is supported by the Fundamental Research Funds for the Central Universities. This work was performed in part at Aspen Center for Physics, which is supported by National Science Foundation grant PHY-1607611. This work was partially supported by a grant from the Simons Foundation.
\end{acknowledgments}

\software{\texttt{COLOSSUS} \citep{Diemer:2017bwl}, \texttt{Astropy} \citep{Robitaille:2013mpa, Price-Whelan:2018hus}. Our main results can be reproduced using the code available at~\url{https://github.com/valerivardanyan/PBH-NS-Mergers}.}

\bibliography{bibliography}{}
\bibliographystyle{aasjournal}

\end{document}